\documentclass[preprint,showkeys, preprintnumbers,amssymb]{revtex4}

\usepackage{graphicx}
\graphicspath{{figs/}}
\begin{document}
\title{A nonequilibrium Green's function study of thermoelectric properties in single-walled carbon nanotubes}
\author{Jin-Wu~Jiang}
\thanks{Electronic mail: phyjj@nus.edu.sg}
    \affiliation{Department of Physics and Centre for Computational Science and Engineering,
             National University of Singapore, Singapore 117542, Republic of Singapore }
\author{Jian-Sheng~Wang}
    \affiliation{Department of Physics and Centre for Computational Science and Engineering,
                 National University of Singapore, Singapore 117542, Republic of Singapore }
\author{Baowen~Li}
        \affiliation{Department of Physics and Centre for Computational Science and Engineering,
                     National University of Singapore, Singapore 117542, Republic of Singapore }
        \affiliation{NUS Graduate School for Integrative Sciences and Engineering,
                     Singapore 117456, Republic of Singapore}
\date{\today}
\begin{abstract}
The phonon and electron transport in single-walled carbon nanotubes (SWCNT) are investigated using the nonequilibrium Green's function approach. In zigzag SWCNT ($n$, 0) with $mod(n,3)\not=0$, the thermal conductance is mainly attributed to the phonon transport, while the electron only has few percentage contribution. The maximum value of the figure of merit ($ZT$) is about 0.2 in this type of SWCNT. The $ZT$ is considerably larger in narrower SWCNT because of enhanced Seebeck coefficient. $ZT$ is smaller in the armchair SWCNT, where Seebeck coefficient is small due to zero band gap. It is found that the cluster isotopic doping can reduce the phonon thermal conductance obviously and enhance the value of $ZT$. The uniaxial elongation and compress strain depresses phonons in whole frequency region, leading to the reduction of the phonon thermal conductance in whole temperature range. Interestingly, the elongation strain can affect the phonon transport more seriously than the compress strain, because the high frequency $G$ mode is completely filtered out under elongation strain $\epsilon >0.05$. The strain also has important effect on the subband edges of the electron band structure by smoothing the steps in the electron transmission function. The $ZT$ is decreased by strain as the reduction in the electronic conductance overcomes the reduction in the thermal conductance.
\end{abstract}

\keywords{thermoelectricity, carbon nanotube, isotopic doping effect, strain effect}
\maketitle

\section{introduction}
The thermoelectric performance is characterized by the figure of merit $ZT$.\cite{GoldsmidHJ1964, GoldsmidHJ1986, NolasGS2001} It is defined as $ZT = T\times\frac{G_{e}S^{2}}{\sigma}$,
where $T$ is the temperature, $G_{e}$ is the electronic conductance, $S$ is the Seebeck coefficient
(or thermopower), and $\sigma$ is the thermal conductance.
Higher value of $ZT$ has been pursued for many decades, since larger $ZT$ implies higher efficiency of a thermoelectric device. The ideal efficiency of a thermoelectric device for electricity generation $\eta_{max}$ is given by:
\begin{eqnarray}
\eta_{max} & = & \frac{T_{H}-T_{L}}{T_{H}}\times\left(1-\frac{1+\frac{T_{L}}{T_{H}}}{\sqrt{1+ZT}+\frac{T_{L}}{T_{H}}}\right),
\end{eqnarray}
where $T_{H/L}$ are the high/low temperatures. This formula tells us that high $ZT$ yields large value
of $\eta_{max}$. It can be employed to estimate the value of $\eta_{max}$ with $T_{L}/T_{H}=0$, which actually sets an upper limit for $\eta_{max}$. It is easy to see that
$\eta_{max}=0.29$ if $ZT=1$, which is about the level of current bismuth based thermoelectric devices;
$\eta_{max}=0.5$, if $ZT=3$, which is able to compete with conventional refrigerators or generators.\cite{MajumdarA}

Although great efforts have been tried, the value of $ZT$ has been on the order of 1.0 for long time.\cite{ChenG, NolasGS, DresselhausM, MahanGD1997} The basic challenge lies in the relationship between the three physical quantities in the above formula for $ZT$. It is quite difficult to modify one quantity independently and keep the other quantities unaffected. The traditional thermoelectric materials are based on the chemical compound of bismuth and tellurium (eg. Bi$_{2}$Te$_{3}$), with the $ZT$ between 0.8 to 1.0.\cite{ChungDY1997, ChungDY2000} They already gain some applications in fields such as portable refrigerators and electric component coolers.
After the year 2004, some progresses have been made in these compound materials, profiting mainly from the development of experimental synthesis techniques. It was found that AgPb$_{m}$SbTe$_{2+m}$ with $m=8$, 10, the n-type semiconductor can obtain a high value of $ZT$ as 2.2 at 800 K under proper doping.\cite{HsuKF} The binary crystalline n-type material, In$_{4}$Se$_{3-\delta}$, can achieve the ZT value of 1.48 at 705 K, which results from a Peierls distortion induced charge density wave instability.\cite{RhyeeJS} Besides the bulk materials, the nano-materials such as bismuth nanowires\cite{HasegawaY} were also considered as possible thermoelectric materials. The p-type nanocrystalline BiSbTe bulk array gives a $ZT$ as high as 1.2 at room temperature and 0.8 at 250$^{o}$C.\cite{PoudelB} In a very recent experiment, the atomic-layer engineering of bismuth telluride is applied to enhance the thermoelectric figure of merit.\cite{TeweldebrhanD}

Besides the traditional Bismuth compound thermoelectric materials, Hicks {\it et~al.} proposed a different direction to obtain high performance thermoelectric materials.\cite{HicksLD19931} It was suggested that the thermoelectric performance can be improved by applying the quantum-well superlattice structures, which are poor thermal conductors\cite{HyldgaardP, ChenG1998, VenkatasubramanianR2000, LinYM, YangN} due to the confinement effect\cite{Balandin1998}. This theoretical proposal was then realized experimentally by the same group,\cite{HicksLD1996} and was further confirmed in other experiments, such as Bi$_{2}$Te$_{3}$/Sb$_{2}$Te$_{3}$ thin-film superlattice,\cite{VenkatasubramanianR}, or embedded PbSeTe quantum dot superlattice.\cite{HarmanTC}

Another direction to achieve high performance thermoelectric materials was predicted by Hicks and Dresselhaus in 1993.\cite{HicksLD19932} In that paper, they suggested that the value of $ZT$ can be greatly enhanced by preparing materials in the form of one-dimensional conductors or nanowires.
In recent years, there have been a great development in the nanotechnology field. Particularly, high quality silicon nanowires with diameters in nanometer size can be synthesized.\cite{Morales, Holmes, WuY} It is a perfect candidate for testing the theoretical suggestion by Hicks and Dresselhaus. In 2009, Boukai {\it et~al.} found that the single component Si shows efficient thermoelectric performance, by changing the diameter of the silicon nanowire among 10 to 20 nm and impurity doping levels.\cite{BoukaiAI} $ZT$ can reach 1.0 at 200 K in the experiment, where the improvement is due to the decrease of the phonon thermal conductance. Another experiment discovered that the $ZT$ can be further enhanced by introducing roughness in the surface of the silicon nanowires,\cite{HochbaumAI} where the thermal conductance was greatly reduced while the electron conductance remains almost unchanged. This experiment displays that the rough silicon nanowires with diameters about 50 nm are poor thermal conductor, while the Seebeck coefficient and electron conductance are similar as those in the doped bulk Si.

After these experiments on the thermoelectric properties of silicon nanowires, some theoretical works have been carried out to explain the increase of $ZT$ in silicon nanowires\cite{VoTTM, ChenX, MarkussenTPRB, MartinP, HochbaumAI, DonadioD, MarkussenT}, or propose some other possible ways to obtain higher $ZT$. In Ref.~\onlinecite{MartinP}, the surface roughness is described by the root-mean-square roughness height and autocovariance length and the obtained thermal conductivity is inverse quadratic proportional to the roughness height. A very small value of thermal conductivity (about 2 W/mK) at room temperature is obtained for 50 nm silicon nanowires, which explains the enhancement of $ZT$ in the experiment.\cite{HochbaumAI} The importance of the amorphous surface on the thermal conductivity is investigated in Ref.~\onlinecite{DonadioD}, where the thermal conductivity was varied in a large range by changing the surface structure. The thermal conductivity is about 100 times smaller than the bulk Si in silicon nanowires with amorphous surface, because of the existence of nonpropagating phonon modes and the decreased lifetime of the propagating modes. In Ref.~\onlinecite{MarkussenT}, Markussen {\it et~al.} proposed theoretically two surface decoration methods to increase the $ZT$ for the silicon nanowire of small diameters. These surface-decorated silicon nanowires have very small thermal conductance, while the electronic conductance are kept considerable large, leading to high $ZT$. The geometry and size effects on the thermoelectric properties in silicon nanowires are also investigated from the electron band structure\cite{ShiLH, LiangG}, or finite element simulation plus analytic modeling\cite{ZhangG}. Some other Si based structures were also proposed to be efficient thermoelectric materials.\cite{LeeJH}

Another attractive nano-material, the carbon-based nanoribbons also received some studies on its thermoelectric properties.\cite{HwangEH, NiX, OuyangY} In Ref.~\onlinecite{NiX}, the nonequilibrium Green's function (NEGF) method combined with the first-principle calculation was applied to study the $ZT$ in graphane (graphene terminated by hydrogen) nanoribbon. $ZT$ was remarkably improved by randomly removing some hydrogen atoms. The thermoelectric performance of the graphene nanoribbon is investigated by Ouyang and Guo.\cite{OuyangY} They solved the atomic phonon and electron transport equations in the NEGF formalism in the ballistic transport regime. The electron band structure was described in the single-$\pi$ orbital tight-binding scheme and the phonons were characterized by the spring mass model. It was found that the quasi-one-dimensional geometry is very important in determining the thermoelectric properties in graphene nanoribbon, and the edge roughness and lattice vacancy reduce the $ZT$ as the electronic conductance is decreased more than the thermal conductance.

Related to the thermoelectric properties, there are two effects in the nanodevices that have attracted much attention. Recently, the size of the cluster can be experimentally controlled in the cluster isotopic doping (CID).\cite{LiX} This particular type of isotopic doping was theoretically studied on its effect on the thermal conductance by Mingo {\it et~al.}\cite{MingoN}. Another important engineering technique is the strain,\cite{Ni, Mohiuddin, Huang, XuY} which can be well controlled experimentally.\cite{BaoW} As the experimental techniques for the isotopic doping and strain have become applicable, it is meaningful to investigate their effects on the thermoelectric properties in the single-walled carbon nanotubes (SWCNT). This forms part of the aim of present paper.

In this paper, we investigate the thermoelectric properties of SWCNT by using the NEGF method in the ballistic transport regime. The lattice dynamics of the phonon is described by the Brenner potential, and the electron band structure is captured by the single-$\pi$ orbital tight-binding scheme. For pure and perfect SWCNT, we study the size and chiral symmetry effect. We find that thermal conductance in the SWCNT ($n$, 0) with $mod(n,3)\not=0$ is mainly determined by the phonon, while the electron thermal conductance is only about few percentage of the total thermal conductance. For armchair SWCNT, the value of $ZT$ is much smaller because of small Seebeck coefficient due to zero band gap in the electron band structure. We then study the CID and strain effect on the thermoelectric properties in the SWCNT. Different from the random isotopic doping, the CID plays a role like a big doping molecule in case of small cluster size, and like an interface in case of large cluster size. The CID reduces the phonon thermal conductance, thus enhancing the $ZT$. For the strain effect, we compare the elongation and compress strain and find that the elongation strain reduces the thermal conductance more efficiently than the compress strain. We also find that the strain considerably smooth the electron conductance at band edges. Finally, the $ZT$ is reduced by strain, because the reduction in the electronic conductance overcomes the reduction in the thermal conductance.

The rest of present paper is organized as follows. In Sec.II, we present the NEGF formulas for both phonon and electron ballistic transport. Sec. III~A is about some calculation details. The calculation results and discussions are given in the Sec. III~B-F. The conclusion is in Sec. V.

\section{NEGF approach}
In the NEGF formalism, a whole system is divided into three different parts (see Fig.~\ref{fig_cfg}): left lead (L), center region (C), and right lead (R).\cite{DattaS, HaugH} We are interested in the finite center region, and treat the two periodic semi-infinite leads as the electron or heat baths. The leads show their effects on the center region through the so-called self-energy as shown in following. The NEGF approach initially found its application in the treatment of the electron transport, then it was borrowed by the phonon transport community.\cite{OzpineciA, CiraciS, YamamotoT, MingoN2006, WangJS2007, WangJS2008} It turns out that the NEGF method can handle the electron and phonon transport in a parallel way, with only few substitutions. Although the NEGF roots in the electron transport filed, in the following we would like to begin with its application in the phonon transport, followed by its usage in the electron field. We focus on the ballistic transport, so the NEGF method can give the exact quantum mechanical results without any approximation. However, the weak electron-phonon scattering\cite{GunlyckeD,OuyangY} and the phonon-phonon scattering are ignored in the ballistic transport. Actually, the phonon-phonon scattering will become important in high temperature region, so the ballistic results overestimate the phonon thermal conductance at high temperatures.

\subsection{NEGF for phonon thermal transport}
The application of NEGF method in the phonon transport can be realized in following five steps.\cite{WangJS2008, WangJSweb} We follow the notations in Ref.~\onlinecite{WangJS2008}. That review paper also contains the definition of different versions of Green's function. More detailed derivations can also be found in Zeng's thesis.\cite{Zeng} 

(1). Write down the Hamiltonian of the whole system, including two leads and the center region:
\begin{eqnarray}
H & = & \sum_{\alpha=L,C,R}H_{\alpha}+\left(u^{L}\right)^{\top}V^{LC}u^{C}+\left(u^{C}\right)^{\top}V^{CR}u^{R},\nonumber\\
\end{eqnarray}
where $u$ is row vector constituted by the vibrational displacement of each atom multiplied by its mass. $V^{LC}$ and $V^{CR}$ are the coupling between leads and center. $H_{\alpha}$ is the linear Hamiltonian for three parts:
\begin{eqnarray}
H_{\alpha} & = & \frac{1}{2}\left(\dot{u}^{\alpha}\right)^{\top}\dot{u}^{\alpha}+\frac{1}{2}\left(u^{\alpha}\right)^{\top}K^{\alpha}u^{\alpha},
\end{eqnarray}
where $K^{\alpha}$ are force constant matrices.

(2). Calculate the retarded Green's function for the three isolated parts: left lead, center region, and right lead:
\begin{eqnarray}
g_{\alpha}^{r} & = & \left[\left(\omega+i\eta\right)^{2}I-K^{\alpha}\right]^{-1}.
\end{eqnarray}
$g_{C}^{r}$ is the retarded
Green's function for the isolated center region, and can be directly calculated from this formula. $g_{L/R}^{r}$ are called the surface Green's function for the two isolated leads. As we will see in following, the calculation of surface Green's function is the most important task in the NEGF approach. But it is not an easy job to obtain the surface Green's function directly from above formula, as leads are semi-infinite. For the simple one-dimensional chain, the surface Green's function have analytical expression.\cite{WangJS2007} But for real systems, it is only possible to do the calculation numerically. Considering the periodic property of the lead, different numerical methods have been proposed to obtain the surface Green's function. The comparison between different methods can be found in Ref.~\onlinecite{WangJS2008}. A direct iterative method is simple yet slow, which increases the lead only one more unit cell after each step.\cite{VelevJ} Lopez Sancho {\it et~al.} developed a fast and stable method, which accounts the effect of $2^{i}$ unit cells for the lead after $i$ interative steps.\cite{SanchoMPL} The most accurate and fast methods is to solve a generalized eigenvalue problem.\cite{LeeDH, UmerskiA, SanvitoS, KrstiPS} However, this method is not so stable in long wave limit. On balance, we accept the second method developed by Lopez Sancho {\it et~al.} in our calculation.

(3). After obtaining the surface Green's function of the leads, we can calculate the retarded self-energy of the leads by:
\begin{eqnarray}
\Sigma_{\alpha}^{r} & = & V^{C\alpha}g_{\alpha}^{r}V^{\alpha C},
\end{eqnarray}
which carries the coupling information between leads and center region. Then calculate the $\Gamma$ function:
\begin{eqnarray}
\Gamma_{\alpha} & = & i\left(\Sigma_{\alpha}^{r}-\Sigma_{\alpha}^{a}\right)=-2Im\Sigma_{\alpha}^{r},
\end{eqnarray}
where the relation $\Sigma_{\alpha}^{a} = \left(\Sigma_{\alpha}^{r}\right)^{\dagger}$ is used.

Calculate the retarded Green's function for the center region
connected with leads:
\begin{eqnarray}
G^{r} & = & \left[\left(\omega+i\eta\right)^{2}I-K^{C}-\Sigma_{L}^{r}-\Sigma_{R}^{r}\right]^{-1}.
\end{eqnarray}

(4). The transmission for phonon is given by the Caroli formula:
\begin{eqnarray}
T[\omega] & = & Tr\left(G^{r}\Gamma_{L}G^{a}\Gamma_{R}\right),
\end{eqnarray}
where $G^{a}=\left(G^{r}\right)^{\dagger}$ is the advanced Green's
function. This formula gives a real number for the transmission of phonon, because $\Gamma_{L}$ and $\Gamma_{R}$ are both real matrix and $G^{a}=\left(G^{r}\right)^{\dagger}$. Another equivalent, more symmetric expression for transmission is constructed by two terms:\cite{Zeng}
\begin{eqnarray}
T[\omega] & = & \frac{1}{2}Tr\left(G^{r}\Gamma_{L}G^{a}\Gamma_{R}\right)+\frac{1}{2}Tr\left(G^{a}\Gamma_{L}G^{r}\Gamma_{R}\right).
\end{eqnarray}

(5). Finally, the phonon thermal conductance can be obtained by Landau formula: 
\begin{eqnarray}
\sigma_{ph} & = & \frac{1}{2\pi}\int d\omega\hbar\omega T_{ph}[\omega]\left[\frac{\partial n(\omega,T)}{\partial T}\right].
\end{eqnarray}
We note that this formula gives the thermal conductance, not conductivity, because in the ballistic region the thermal conductance does not dependent on the size of the system, while the thermal conductivity will diverge with increasing size. This formula accounts for the ballistic phonon thermal conductance, so it will overestimate the value of thermal conductance at high temperatures. The phonon-phonon scattering becomes important at high temperatures, and will depress the phonon thermal conductance. This nonlinear interaction can be included through a nonlinear self-energy in the NEGF theme, or treated by the Boltzmann transport theory.

\subsection{NEGF for electron transport}
The procedure for NEGF in the electron transport is the same as the phonon transport after two substitutions:
\begin{eqnarray}
K_{C} & \rightarrow & H_{e},\\
\left(\omega+i\eta\right)^{2} & \rightarrow & E+i\eta,
\end{eqnarray}
where the force constant matrix $K_{C}$ matrix is replaced by the electron Hamiltonian
$H_{e}$, and the square of frequency $\omega ^{2}$ is replaced by the energy of
electron $E$.
After the above five steps, electron transmission function is obtained, and can be used to calculate all electronic quantities: electronic conductance $G_{e}$, electron thermal conductance $\sigma_{e}$, and Seebeck coefficient $S$. Formulae for these electronic quantities in the ballistic transport were first obtained in 1986.\cite{Sivan} A straightforward derivation is given in Ref.~\onlinecite{Esfarjani} by Esfarjani {\it et~al.}. Following are the three key steps in the derivation.

(1). In the ballistic transport, the two Landauer-Buttiker formulae for electronic current and electron
thermal current are:
\begin{eqnarray}
I & = & \frac{2q}{h}\int_{-\infty}^{+\infty}dET(E)\left[f(E,\mu_{L})-f(E,\mu_{R})\right],\\
I_{Q} & = & \frac{2}{h}\int_{-\infty}^{+\infty}dET(E)\left[f(E,\mu_{L})-f(E,\mu_{R})\right]\left(E-\mu\right),\nonumber\\
\end{eqnarray}
where $q=-e$ is the charge for electron, and $h$ is the Plank constant. $\mu$ is the chemical potential. $f(E,\mu)$ is the Fermi-Dirac distribution function, which is also temperature dependent. 

(2). Under the linear response approximation, we can obtain the linear coherent transport formalism for different physical quantities. The above two formulae can be expanded in terms of small variables $\Delta\mu$ and $\Delta T$:
\begin{eqnarray}
I & = & \Delta\mu\frac{2q}{h}\int_{-\infty}^{+\infty}dET(E)\frac{\partial f}{\partial\mu}+\Delta T\frac{2q}{h}\int_{-\infty}^{+\infty}dET(E)\frac{\partial f}{\partial T},\nonumber\\
&&\\
I_{Q} & = & \Delta\mu\frac{2}{h}\int_{-\infty}^{+\infty}dET(E)\frac{\partial f}{\partial\mu}\left(E-\mu\right)\nonumber\\
&&+\Delta T\frac{2}{h}\int_{-\infty}^{+\infty}dET(E)\frac{\partial f}{\partial T}\left(E-\mu\right).
\end{eqnarray}

(3). After introducing an intermediate function
\begin{eqnarray}
L_{n}(\mu,T) & = & \frac{2}{h}\int dET_{e}(E)\times\left(E-\mu\right)^{n}\times\left[-\frac{\partial f\left(E,\mu,T\right)}{\partial E}\right],\nonumber\\
\label{eq_Ln}
\end{eqnarray}
the electronic quantities can be calculated following their definitions.

The electronic conductance is:
\begin{eqnarray}
G_{e} & = & -\frac{I}{V}|_{\Delta T=0}
 =  \frac{-1}{V}\times\Delta\mu\frac{2q}{h}\int_{-\infty}^{+\infty}dET(E)\frac{\partial f}{\partial\mu}\nonumber\\
 &=&  q^{2}L_{0}.
\end{eqnarray}

The Peltier coefficient is:
\begin{eqnarray}
\Pi & = & \frac{I_{Q}}{I}|_{\Delta T=0}
 =  \frac{\Delta\mu\frac{2}{h}\int_{-\infty}^{+\infty}dET(E)\frac{\partial f}{\partial\mu}\left(E-\mu\right)}{\Delta\mu\frac{2q}{h}\int_{-\infty}^{+\infty}dET(E)\frac{\partial f}{\partial\mu}}\nonumber\\
& = & \frac{L_{1}}{qL_{0}}.
\end{eqnarray}

The Seebeck coefficient is:
\begin{eqnarray}
S & = & -\frac{V}{\Delta T}|_{I=0}
 = -\frac{1}{e}\times\frac{\frac{2q}{h}\int_{-\infty}^{+\infty}dET(E)\frac{\partial f}{\partial T}}{\frac{2q}{h}\int_{-\infty}^{+\infty}dET(E)\frac{\partial f}{\partial\mu}}\nonumber\\
 &=&  \frac{1}{qT}\times\frac{L_{1}}{L_{0}}.
\label{eq_S}
\end{eqnarray}

The thermal conductance is:
\begin{eqnarray}
\sigma _{e} & = & -\frac{I_{Q}}{\Delta T}|_{I=0}\nonumber\\
& = & -\frac{1}{\Delta T}\times[\Delta\mu\frac{2}{h}\int_{-\infty}^{+\infty}dET(E)\frac{\partial f}{\partial\mu}\left(E-\mu\right)\nonumber\\
&& + \Delta T\frac{2}{h}\int_{-\infty}^{+\infty}dET(E)\frac{\partial f}{\partial T}\left(E-\mu\right)]\nonumber\\
& = & \frac{1}{T}\times\left(L_{2}-\frac{L_{1}^{2}}{L_{0}}\right).
\end{eqnarray}

\section{results and discussion}
\subsection{structure and calculation details}
\begin{figure}[htpb]
  \begin{center}
    \scalebox{1.2}[1.2]{\includegraphics[width=7cm]{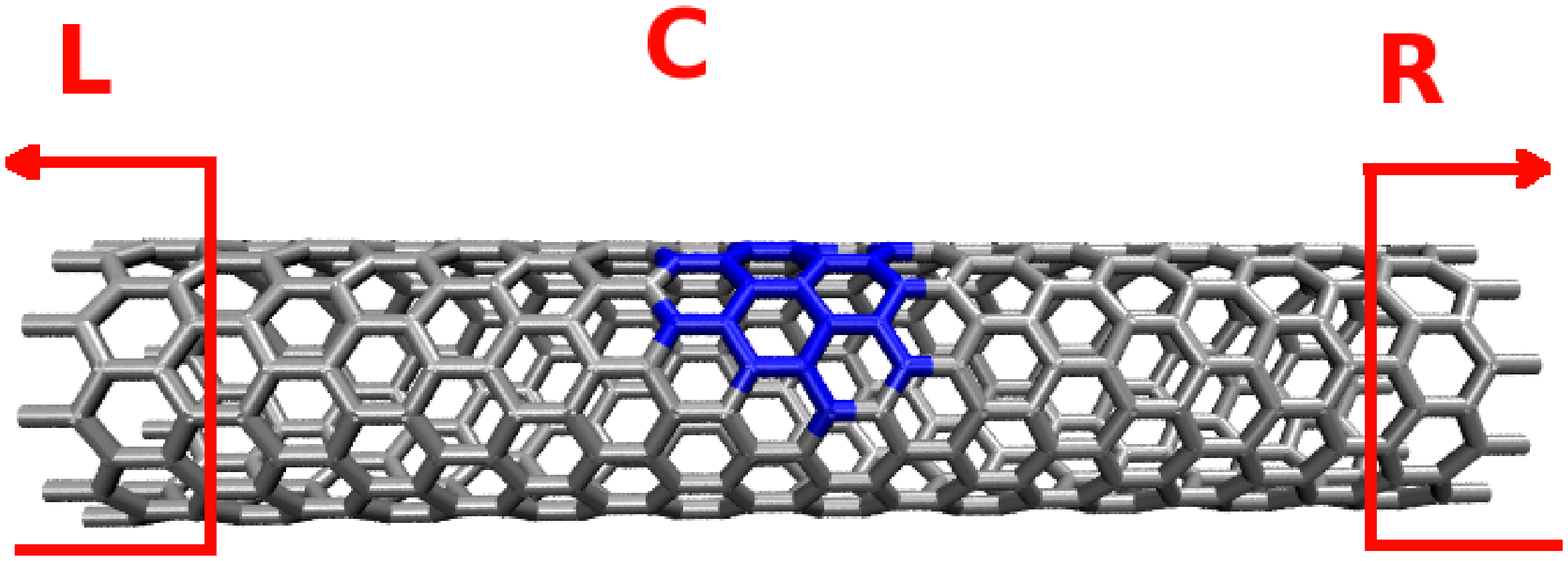}}
  \end{center}
  \caption{(Color online) Configuration of the structure. SWCNT is divided into three regions: left lead, center region, and right lead. In the center region, those atoms in blue online are isotopic doping.}
  \label{fig_cfg}
\end{figure}

Fig.~\ref{fig_cfg} displays the configuration of SWCNT (10, 0). The whole tube is divided into two leads and a center region as required by the NEGF scheme. In each lead, there are two columns. The semi-infinity leads can be generated by periodically repeating these two columns. Each column contains forty carbon atoms, and is large enough to ensure that atoms in each column only interact with atoms in its two neighboring columns within the applied Brenner potential.\cite{Brenner} After these manipulations, the leads in the SWCNT are quasi-one-dimensional systems with first-nearest-neighboring interactions. The surface Green's function of this type of system can be calculated by the efficient iterative method developed by Lopez Sancho {\it et~al.}. Blue atoms in the center region are $^{14}$C isotopic doping, forming a cluster. The size of the cluster, $r_{c}$, is 4.0~{\AA} in the figure.

The lattice dynamical force constant matrices are generated from the Brenner potential\cite{Brenner} implemented in the ``General Utility Lattice Program" (GULP)\cite{Gale}. The GULP is used to optimize the structure of the SWCNT. The equilibrium position under strain can be realized by stretching the SWCNT and then optimizing the structure with both ends fixed. The exported force constant matrix from GULP is further partitioned according to Fig.~\ref{fig_cfg}. Finally the input force constant matrices for the NEGF are obtained, such as $K_{L}$, $K_{R}$, $K_{C}$ and the coupling $V^{LC}$, $V^{CR}$. In the NEGF procedure, leads are regarded as perfect and pure. So the CID and the strain effects are only applied to $K_{C}$ of the center region.

The total Hamiltonian for electron is described by the single-$\pi$ orbital tight-binding scheme\cite{MahanGD}:
\begin{eqnarray}
H & = & \sum_{j\delta\sigma}J_{\delta}\left(\epsilon\right)\left[C_{A,j\sigma}^{\dagger}C_{B,j+\delta\sigma}+C_{B,j+\delta\sigma}^{\dagger}C_{A,j\sigma}\right],
\end{eqnarray}
where $A$ and $B$ are the two nonequivalent carbon atoms, and the summation $\sum_{\delta}$ is taken over the three first-nearest-neighboring bonds. The hopping parameter is generalized to include the effect of the strain ($\vec{\epsilon}$) by treating the strain as a kind of phonon vibration,
\begin{eqnarray}
J_{\delta}\left(\epsilon\right) = J_{0}+J_{1}\hat{\delta}\cdot\vec{\epsilon},
\end{eqnarray}
where the unit vector $\hat{\delta}$ is from atom $(j,A)$ to atom $(j+\delta,B)$. Parameters $J_{0}=3.0$eV, and $J_{1}/J_{0}=2.0$~{\AA} are taken from Ref.~\onlinecite{MahanGD}. There are different optional formulas to describe the strained hopping parameter.\cite{YangL,PereiraVM} To avoid possible arbitrariness for the strained hopping parameter, we generalize the above electron-phonon interaction Hamiltonian to describe the strain effect by treating the strain as a kind of phonon. This is reasonable since the long-wave phonon is actually very similar to a uniform strain. It should be noted that the tight-binding model does not account for carrier renormalization effects when the chemical potential is moved near (or inside) the valence or conduction bands, which plays important role on electronic bands structure of semiconducting SWCNT.

\subsection{pure SWCNT}
\begin{figure}[htpb]
  \begin{center}
    \scalebox{1.0}[1.0]{\includegraphics[width=7cm]{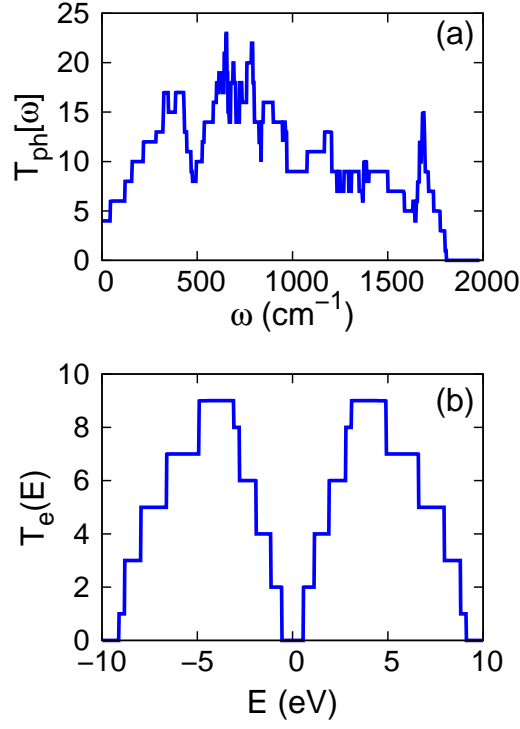}}
  \end{center}
  \caption{(Color online) Transmission function for pure and perfect SWCNT (10, 0). (a). phonon, (b) electron.}
  \label{fig_T_pure}
\end{figure}

We first investigate the thermoelectric properties in the perfect and pure SWCNT (10, 0). Fig.~\ref{fig_T_pure} (a) and (b) show the transmission function for phonon and electron, respectively. They are some regular steps, which simply implies that all phonons or electrons from the lead can pass through the center region without any scattering. There is no decoherence mechanisms here. In the ballistic transport, the NEGF transmission function are essentially the number of available channels for conduction, because the electron-phonon and phonon-phonon scattering are ignored. Another way to get the transmission function is by counting the number of phonon/electron states in their dispersion/bandstructure. However, the NEGF approach can include various terms in the self-energies when higher-order effects (such as decoherence due to electron-phonon and phonon-phonon interactions) are considered. By considering the nonlinear interactions, one major difference from ballistic transport is that the transmission function will depends on temperature, because the phonon-phonon scattering depends on temperature. The high energy phonon modes will have smaller transmission coefficients due to scattering, leading to smaller phonon thermal conductance at high temperatures. This nonlinear effects at high temperatures are out of the scope of present work. We are focusing on the ballistic transport. Panel (a) gives us some information of phonon spectrum in the SWCNT. There are totally four acoustic modes in low frequency region, three of which are the usual translational acoustic modes. The other one is the twisting mode, originating from the rotational invariant cylindrical shell configuration of the SWCNT.\cite{MahanGD2004, JiangJW2006} In the frequency region around 800 cm$^{-1}$, there are mainly the optical vibrational modes in the radial direction. And the peaks in high frequency correspond to the highest in-plane optical phonon modes ($G$ mode). Panel (b) shows a band gap in SWCNT (10, 0). This corresponds to the rule for the SWCNT ($n_{1}$, $n_{2}$), that nonzero band gap occurs in case of $mod(n_{1}-n_{2}, 3)\not=0$.\cite{SaitoR} The value of the band gap (about 1.0 eV) is close to the first-principle calculation\cite{ShahD} and other numerical method for this type of SWCNT.\cite{MencarelliD}

\begin{figure}[htpb]
  \begin{center}
    \scalebox{1.2}[1.2]{\includegraphics[width=7cm]{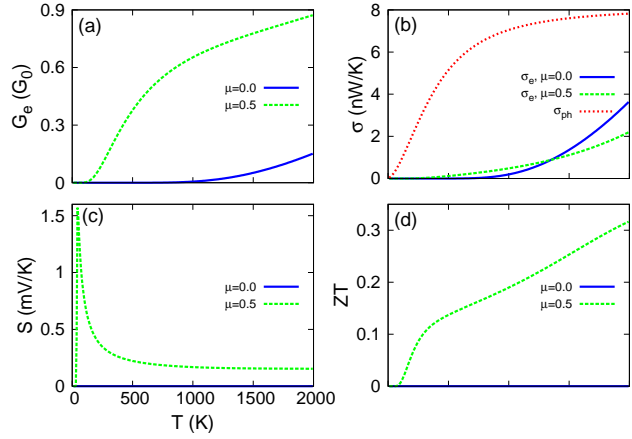}}
  \end{center}
  \caption{(Color online) Temperature dependence for thermoelectric properties in pure and perfect SWCNT (10, 0): (a). electric conductance; (b). electron and phonon thermal conductance; (c). Seebeck coefficient; (d). $ZT$. The chemical potential is in the unit of eV.}
  \label{fig_zt_T}
\end{figure}

Fig.~\ref{fig_zt_T} gives the temperature dependence for different electronic and phonon properties. We have studied the electronic properties at different chemical potentials $\mu=0.0$ and 0.5 eV. The phonon thermal conductance does not depend on the chemical potential. In panel (a), the electronic conductance with $\mu=0$ is almost zero at temperatures up to 1000 K, because of the band gap in the SWCNT (10, 0) and electrons are in the valence band. Only at very high temperature, some electrons can be excited into the conductance band and contribute to the electronic conductance. When $\mu=0.5$ eV, the Fermi surface is moved close to conductance subbands. So the electron can be easily excited into conductance subband and the electronic conductance has reasonably large value around room temperature. Panel (b) distinctly shows that the thermal conductance is mainly contributed by the phonon. The electron thermal conductance is ignorable, especially in case of $\mu=0$. The increase of the chemical potential can enhance the electron thermal conductance at low temperature. These phenomena are similar to that obtained by Saito {\it et~al.} with Landauer formula in the graphene nanoribbon system.\cite{SaitoK} Panel (c) shows that the Seebeck coefficient is zero at $\mu=0$ because of the symmetric electron band structure. At $\mu=0.5$ eV, the curve of $S(T)$ has a peak around 100 K. This peak is actually corresponding to the maximum point of the electron-transmission-weighted average value of $(E-\mu)$ in Eq.~(\ref{eq_Ln}) for $L_{1}$.\cite{OuyangY} Finally, the panel (d) shows the combined effect from all of the above four quantities. At $\mu=0$, the $ZT$ is zero in whole temperature range, as the Seebeck coefficient $S=0$. For $\mu=0.5$ eV, the $ZT$ has zero value in extremely low temperature region, where the electronic conductance $G_{e}$ is almost zero. Below 500 K, the curve increases rapidly, since the $G_{e}$ increases quickly in this temperature range. In the meantime, $S$ has a relative large value and the thermal conductance is not very large yet. Above 500 K, the $ZT$ increases moderately, as the $S$ is now kept as a small constant and the thermal conductance is large.

\begin{figure}[htpb]
  \begin{center}
    \scalebox{1.2}[1.2]{\includegraphics[width=7cm]{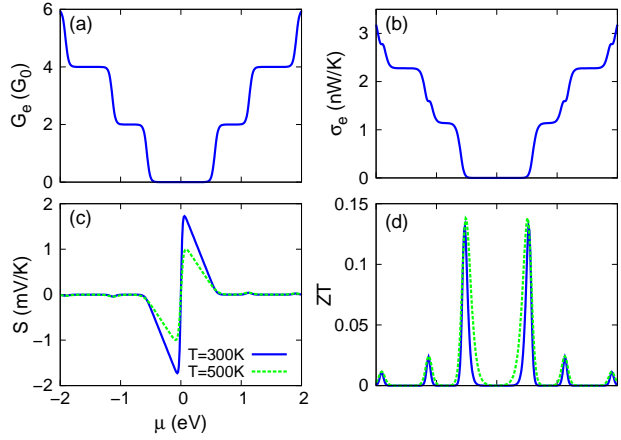}}
  \end{center}
  \caption{(Color online) The thermoelectric properties with different chemical potential for SWCNT (10, 0). (a). electric conductance; (b). electron thermal conductance; (c). Seebeck coefficient; (d). $ZT$.}
  \label{fig_zt_pure}
\end{figure}

Fig.~\ref{fig_zt_pure} is chemical potential dependence for the thermoelectric properties of the perfect and pure SWCNT. The chemical potential $\mu$ is in the range $[-2, 2]$eV. The unit of the electronic conductance is the conductance quantum $G_{0}=2e^{2}/h=77.4\mu$S. In panel (a), if the chemical potential is in the middle of the valence and conductance subbands, very few electrons can be excited to conductance band, thus leading to an almost zero value of the electronic conductance. If the chemical potential moves into the subbands, the conduction channels increase step by step. As a result of the increase of the electron conductance channels, the electron thermal conductance also increases in a step fashion as shown in panel (b). Panel (c) shows the antisymmetric structure of the Seebeck coefficient, because the subband is symmetric, $L_{1}$ is odd function about $(E-\mu)$ and the transmission function is an even function. The $S$ arrives its maximum point around $\mu\approx k_{B}T$. The peak at 500 K is lower than 300 K, because the Fermi-Dirac distribution function $f(E,\mu)$ is smoother at the Fermi surface at higher temperature. As a result, the value of $\partial f/\partial E$ in Eq.~(\ref{eq_Ln}) for $L_{1}$ is smaller at higher temperature, leading to smaller value of $S$ from Eq.~(\ref{eq_S}). Panel (d) is the $ZT$, combining all effects from the above three quantities. $ZT$ has peaks at the edge of each subbands, because the conduction channel increases suddenly when the chemical potential arrive at the edge of subbands. The first two have the largest value among all peaks, due to the large value of $S$ inside this chemical potential region and Seebeck coefficient presents as $S^{2}$ in the formula of $ZT$. The peak in $ZT$ curve is sharper at lower temperature, because of sharper step in the partial function $\partial f/\partial E$.

\subsection{size effect}
\begin{figure}[htpb]
  \begin{center}
    \scalebox{1.0}[1.0]{\includegraphics[width=7cm]{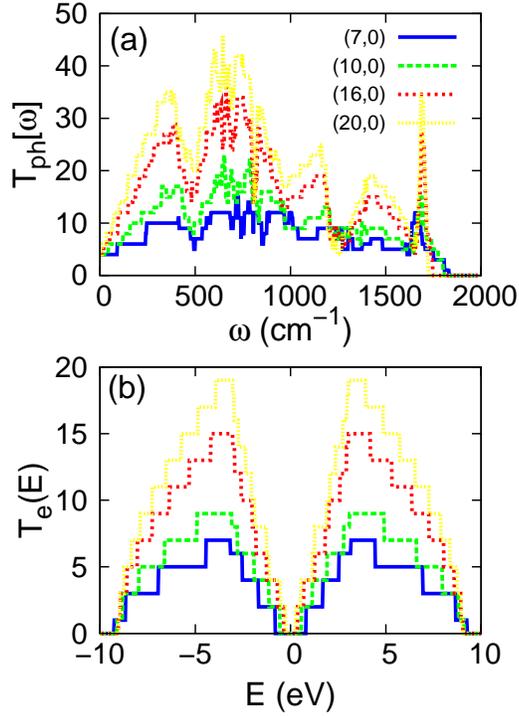}}
  \end{center}
  \caption{(Color online) Phonon and electron transmission function in zigzag SWCNT with different radius: (a) for phonon and (b) for electron.}
  \label{fig_T_radius}
\end{figure}

\begin{figure}[htpb]
  \begin{center}
    \scalebox{1.1}[1.1]{\includegraphics[width=7cm]{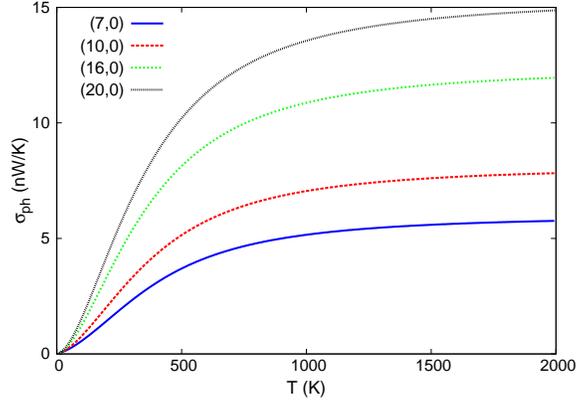}}
  \end{center}
  \caption{(Color online) Phonon thermal conductance in zigzag SWCNT with different radius. Inset shows the $\sigma_{ph}/radius$ v.s temperature.}
  \label{fig_K_p_radius}
\end{figure}

We now study the thermoelectric properties of perfect and pure SWCNT ($n$, 0) with different radius. Tubes compared here are in the same type with $mod(n_{1}-n_{2}, 3)\not=0$. Fig.~\ref{fig_T_radius}.~(a) shows that the transport channel for the phonon increases proportionally with the increase of radius. Panel (b) shows similar increasing behavior in the transmission function for the electron transport. We note another important phenomenon in the electron transmission function; i.e the band gap becomes narrower in thicker SWCNT with smaller curvature. Actually, the band gap turns to disappear in the large radius limit, which is the result of graphene sheet. Fig.~\ref{fig_K_p_radius} compares the phonon thermal conductance between different SWCNT. The phonon thermal conductance increases with increasing radius, while the value of $\sigma_{ph}/radius$ is almost independent of radius as shown by the inset. We can see that the phonon thermal conductance reaches a platform in high temperature region where all phonon modes are excited. This is the result of the ballistic transport. If the phonon-phonon scattering is considered, the curve will show decreasing behavior at high temperature due to finite lifetime of phonons. Similar phenomena can be seen in following figures about the temperature dependence of the phonon thermal conductance.

\begin{figure}[htpb]
  \begin{center}
    \scalebox{1.2}[1.2]{\includegraphics[width=7cm]{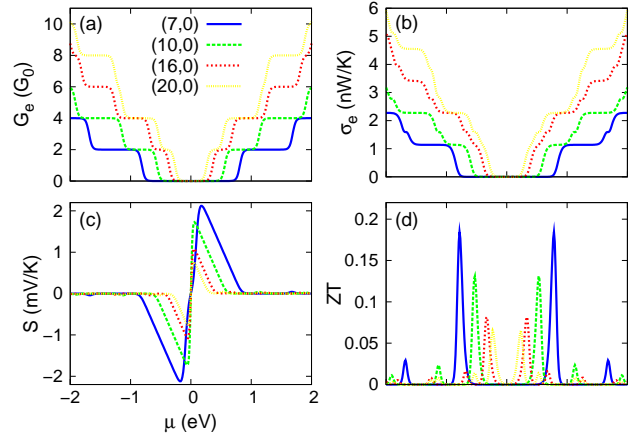}}
  \end{center}
  \caption{(Color online) Thermoelectric properties in zigzag SWCNT at 300 K with different radius. (a). electric conductance; (b). electron thermal conductance; (c). Seebeck coefficient; (d). $ZT$.}
  \label{fig_zt_radius}
\end{figure}

Corresponding to the decrease of band gap, the electronic conductance and electron thermal conductance also have smaller gap as shown in Fig.~\ref{fig_zt_radius}~(a) and (b). For the same reason, panel (c) shows that the Seebeck coefficient has smaller value in larger SWCNT. As a result, the $ZT$ in thicker SWCNT is smaller. The two peaks in the $ZT$ curve also get closer to each other as the radius increase due to the decrease of band gap.

In the ballistic transport regime, the electron or phonon transport should not depend on the length of the system, because all electron or phonons from baths can pass through the center region perfectly. We have checked that this is indeed the case, indicating the correctness of our approach.

\subsection{chiral symmetry effect}
\begin{figure}[htpb]
  \begin{center}
    \scalebox{1.0}[1.0]{\includegraphics[width=7cm]{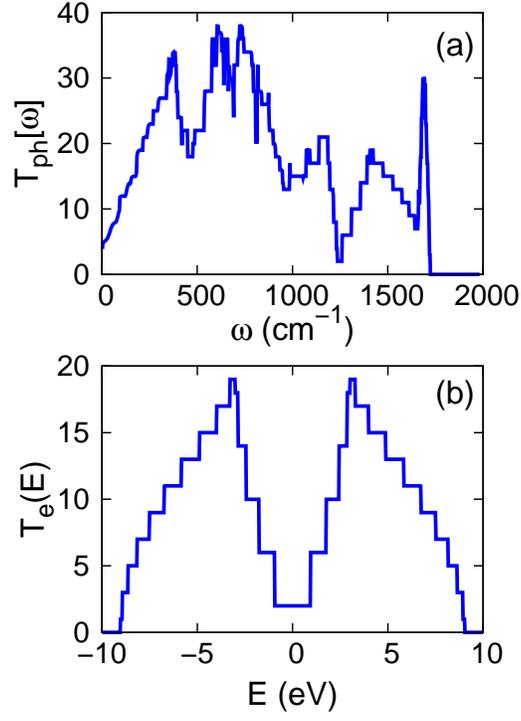}}
  \end{center}
  \caption{(Color online) Phonon transmission function in (a), and electron transmission function in (b) for the armchair SWCNT (10, 10).}
  \label{fig_T_arm}
\end{figure}

\begin{figure}[htpb]
  \begin{center}
    \scalebox{1.1}[1.1]{\includegraphics[width=7cm]{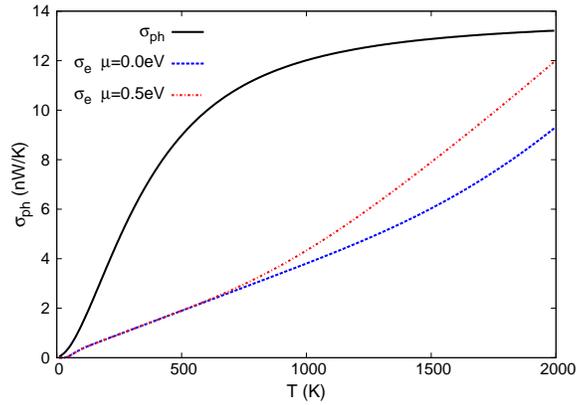}}
  \end{center}
  \caption{(Color online) Phonon thermal conductance in armchair SWCNT (10, 10).}
  \label{fig_K_p_arm}
\end{figure}

To see the chiral symmetry effect, we study the armchair SWCNT ($n$, $n$) fulfilling the condition $mod(n_{1}-n_{2}, 3)=0$. Band gap is zero in this type of SWCNT. Fig.~\ref{fig_T_arm}.~(a) shows the phonon transmission function in armchair SWCNT, where the number of phonons around 1300 cm$^{-1}$ is quite small compared with zigzag SWCNT. The electron transmission function in (b) confirms zero band gap in armchair SWCNT. The phonon/electron thermal conductance of this armchair SWCNT is given in Fig.~\ref{fig_K_p_arm}. In SWCNT (10, 10), the electron has significant contribution to the thermal conductance, which results from its metallic property. Especially at higher chemical potential, the electron thermal conductance can be larger than the phonon thermal conductance at high temperatures.

\begin{figure}[htpb]
  \begin{center}
    \scalebox{1.2}[1.2]{\includegraphics[width=7cm]{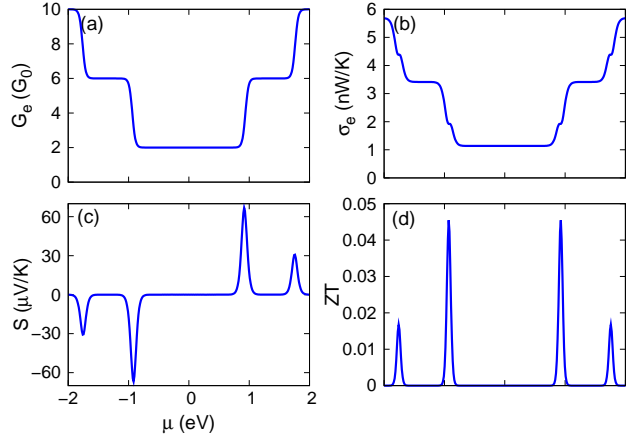}}
  \end{center}
  \caption{(Color online) Thermoelectric properties at 300 K in SWCNT (10, 10). (a). electric conductance; (b). electron thermal conductance; (c). Seebeck coefficient; (d). $ZT$.}
  \label{fig_zt_arm}
\end{figure}

Fig.~\ref{fig_zt_arm}.~(a) and (b) show that the electronic conductance and electron thermal conductance has considerably large value even with $\mu=0$, resulting from zero band gap in the electron band structure. We note that the extra tiny steps in (b) are the results of the competition between the three intermediate functions $L_{0}$, $L_{1}$ and $L_{2}$ involved in the formula of electronic conductance. Panel (c) shows that the Seebeck coefficient in armchair SWCNT is much smaller than that of the zigzag SWCNT. This difference indicates the importance of the band gap for large Seebeck coefficient. In semiconductor with band gap, electrons at the conduction subband edge have major contribution to the electron transport. From Eq.~(\ref{eq_Ln}), these electrons have larger value of $E-\mu$, giving larger value of $L_{1}$. As a result, the Seebeck coefficient from Eq.~(\ref{eq_S}) is large.\cite{OuyangY} Similarly, the Seebeck coefficient in graphene (zero band gap) is also very small (in the order of $\mu$S). As a result of very small Seebeck coefficient, the armchair SWCNT has very small $ZT$ as shown in panel (d).

\subsection{cluster isotopic doping effect}
\begin{figure}[htpb]
  \begin{center}
    \scalebox{1.0}[1.0]{\includegraphics[width=7cm]{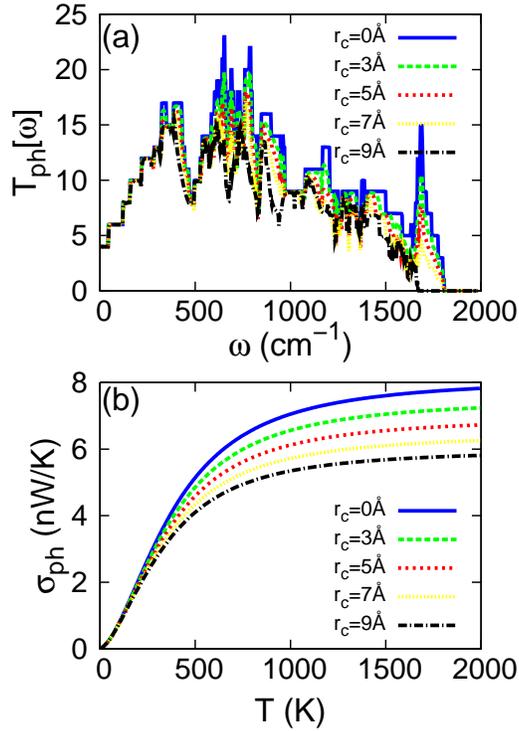}}
  \end{center}
  \caption{(Color online) The isotopic cluster doping effect on the phonon transport for SWCNT (10, 0): (a). transmission function, and (b). phonon thermal conudctance.}
  \label{fig_T_p_rc}
\end{figure}

\begin{figure}[htpb]
  \begin{center}
    \scalebox{1.2}[1.2]{\includegraphics[width=7cm]{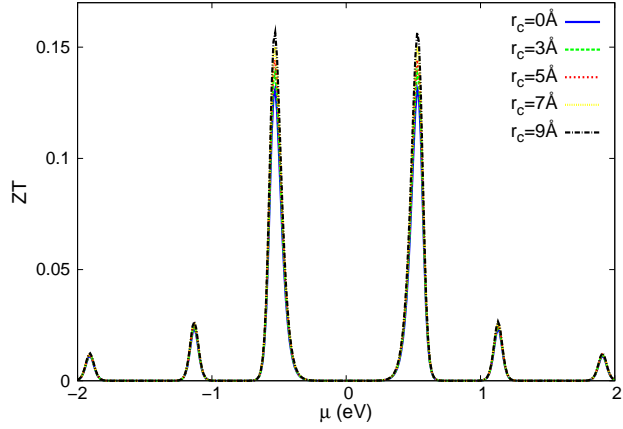}}
  \end{center}
  \caption{(Color online) $ZT$ value at 300 K with different isotopic cluster doping size for SWCNT (10, 0).}
  \label{fig_zt_rc}
\end{figure}

Now we consider the CID effect on the thermoelectric properties of SWCNT (10, 0). We consider a pure SWCNT of $^{12}$C doped by isotopic $^{14}$C. This is a new type of isotopic doping, different from the usual random doping process. In the CID, the doping atoms are constrained within an area with radius $r_{c}$. If $r_{c}$ is small, CID is actually a big doping molecule. If the radius $r_{c}$ is very large, it acts like an interface. The CID can affect the phonon thermal conductance, yet has no effect on the electron transport properties. Fig.~\ref{fig_T_p_rc} is the phonon thermal properties at different cluster size $r_{c}$. Panel (a) shows that CID has limited effect on the low frequency phonon, as the CID plays like one molecule for small $r_{c}$. So the low frequency phonon still has large transmission probability. If the radius $r_{c}$ is large, now the ICD looks like an interface, which is easier for the low frequency phonon to pass through.\cite{ChenKQ2000,ChenKQ2002} Actually, isotopes would scatter phonons of wavelength of the same order as their radius. However, the size of the isotope is in the order of [0, 9] angstrom. This is much lesser than wavelength of the long wavelength phonons, so the low frequency phonon is easier to pass through. The most important effect of the CID is in the frequency range around 800 cm$^{-1}$, and the highest frequency region around 1600 cm$^{-1}$. These two frequency regions are mainly for the optical phonon modes in the three directions. For $r_{c}=9$~{\AA}, the high frequency phonons are filtered out completely. With the increase of the CID size, the phonon thermal conductance decreases gradually. This is different from what have been observed in the random isotopic doping, where the thermal conductivity decrease rapidly in low concentrate region and saturate in high concentrate, because the localized modes plays an important role in the random isotopic doping case, while the CID is more or less like an interface with large $r_{c}$. Fig.~\ref{fig_zt_rc} shows the $ZT$ at different CID size. With the increase of $r_{c}$, the value of $ZT$ will increase. The CID can only enhance the value of $ZT$ slightly, as the only effect from CID is the reduction of thermal conductance which is actually small. The position of the peak in the $ZT$ figure does not affected by the CID, as the main feature of the $ZT$ is determined by the electronic contribution instead of the phonon thermal conductance.

\subsection{strain effect}
\begin{figure}[htpb]
  \begin{center}
    \scalebox{1.0}[1.0]{\includegraphics[width=7cm]{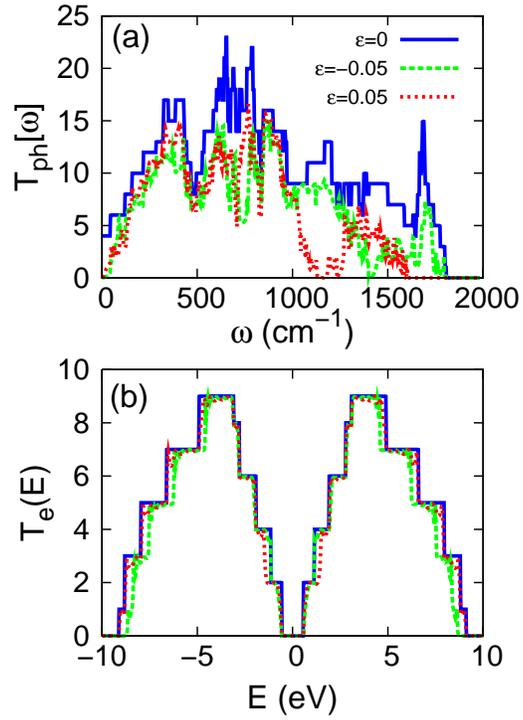}}
  \end{center}
  \caption{(Color online) Strain effect on the phonon transmission function (a) and electron transmission function (b) for SWCNT (10, 0).}
  \label{fig_T_strain}
\end{figure}

Fig.~\ref{fig_T_strain} demonstrates the effect of the strain on the phonon and electron transmission functions. Both elongation and compress strains are studied. Panel (a) shows that the strain favors to filter out the low frequency phonons. The situation has become quite difficult for low frequency phonons to get through the center region even under a very small strain. This is quite comprehensible, because the strain can be regarded as a kind of long wavelength acoustic phonon mode. It prefers to resonate with the low frequency acoustic phonons in the SWCNT, and filter them out efficiently. The figure shows that in case of $\epsilon >0$, phonons around 1200 cm$^{-1}$ are almost filtered out completely, and the $G$ mode shows a red shift. The $G$ mode is the in-plane optical phonon modes in SWCNT, and this red shift has been observed in the experiment.\cite{Mohiuddin, Huang} The $G$ mode will also be filtered out under larger elongation strain, eg. $\epsilon=0.05$ in the figure. If $\epsilon <0$, phonons around 1400 cm$^{-1}$ will be filtered out and the $G$ mode shows blue shift. The suppress of phonon modes at 1200 cm$^{-1}$ or 1400 cm$^{-1}$ is related to the vibrational properties of these phonons which are determined by the interactions in the SWCNT. The interaction is described by the Brenner potential, where the interaction range is upto second-nearest neighboring atoms. The phonon modes at 1200 (1400) cm$^{-1}$ corresponds to the optical phonons at high-symmetric K (M) point at the edge of the Brillouin zone of graphene.\cite{Maultzsch,Mohr} They are not Raman active modes in pure graphene. However, they can lead to the so-called Raman D peak if graphene has some defects.\cite{Ferrari} It means that these two modes are more sensitive to the defect of the structure. As a result, they show more obvious response to the strain, i.e suppressed by strain. Our results indicate that the elongation strain can cause obvious modification to the K point, while the compress strain has significant influence on the M point. Panel (b) displays the effect of the strain on the electron transmission function. The strain smooths the steps in the electron transmission function. This is because we have treated the strain as a kind of phonon vibration, thus the strain effect can be considered from the electron-phonon interaction Hamiltonian. In this Hamiltonian, the effect of the strain is to modify the hopping parameter of the tight-binding model, which will shift and deform the electron bands. However, because of the simplicity of the electron bands from tight-binding model, the shift and deformation do not change the number of the bands in most energy region, except those at the band edge. In the ballistic transport regime, the transmission function is determined by the number of electron bands. So the transmission function and other electronic properties are only affected by strain through smoothing out the band edges. The shift and deformation of the electron bands will be accumulated from low energy region to higher energy region. As a result, the strain effect will be more obvious for higher energy sub-bands, which leads to more asymmetry between positive and negative strain for higher sub-bands.

\begin{figure}[htpb]
  \begin{center}
    \scalebox{1.2}[1.2]{\includegraphics[width=7cm]{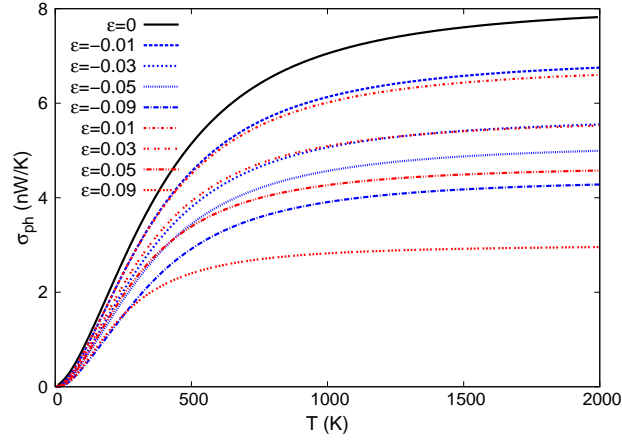}}
  \end{center}
  \caption{(Color online) Strain effect on the phonon thermal conductance for SWCNT (10, 0).}
  \label{fig_K_p_strain}
\end{figure}

Fig.~\ref{fig_K_p_strain} shows strain can significantly reduce the phonon thermal conductance. The reduction is still obvious even in low temperature region, because the strain can filter out both low frequency phonons and the high frequency phonons. This figure also shows that the elongation strain can affect the phonon thermal conductance more distinctly than the compress strain. For example, the $\sigma_{ph}$ is reduced by about 70$\%$ at high temperature under $\epsilon=0.09$, while this value is about 50$\%$ under $\epsilon=-0.09$. This is because the $G$ mode in the SWCNT shows red shift under small elongation strain, and will be filtered out completely under larger elongation strain. This effect sheds some light on the nonequivalent of the elongation and compress strain in the SWCNT.

\begin{figure}[htpb]
  \begin{center}
    \scalebox{1.2}[1.2]{\includegraphics[width=7cm]{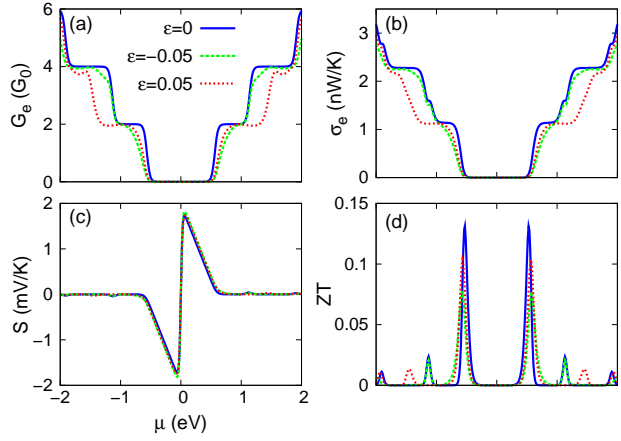}}
  \end{center}
  \caption{(Color online) Strain effect on the thermoelectric properties at 300 K for SWCNT (10, 0): (a). electric conductance; (b). electron thermal conductance; (c). Seebeck coefficient; (d). $ZT$.}
  \label{fig_zt_strain}
\end{figure}

Fig.~\ref{fig_zt_strain} shows strain effect on the electron transport properties and the $ZT$. The electronic conductance will have a smoother step around these subband edges, because the transmission function has been smoothed at the subband edge. Correspondingly, the electron thermal conductance has similar changes in its curve as shown in panel (b). Panel (c) shows that the Seebeck coefficient is almost unaffected by the strain, because the strain only affect the subband edge structure, which actually has limited effect on the Seebeck coefficient. Panel (d) displays the effect of strain on the value of $ZT$. The value of $ZT$ decreases under strain, as the peak of the $ZT$ is due to the sudden jump in the electron transport channels and this jump has now been smoothed out by the strain. We have checked that the phonon and electron properties do not depend on the length of the SWCNT if the same strain is applied, because the phonon and electron are both described by short-range interaction and the applied strain is uniform.

\section{conclusion}
To conclude, the thermoelectric properties of SWCNT are systematically studied in the NEGF scheme in the ballistic transport regime. For the pure and perfect SWCNT, we study the size and chiral symmetry effect. It was found that in the semi-conductor SWCNT the thermal conductance is mainly determined by the phonon, while the electron has very limited contribution.  The maximum value of $ZT$ is about 0.2 in the semi-conductor SWCNT. For metallic SWCNT, the value of the $ZT$ is much smaller, because of the small value of the Seebeck coefficient due to zero band gap in the electron band structure.

Besides the pure and perfect SWCNT, we also study the thermoelectric properties in the SWCNT after CID or under strain. The CID acts like a big doping molecule in case of small cluster size, and like an interface in case of large cluster size. This type of isotopic doping can reduce the phonon thermal conductance, thus enhancing the $ZT$. We comparatively study the effect of elongation and compress strain, and find that the elongation strain can reduce the thermal conductance more efficiently than the compress strain. The strain also considerably modifies the curve of electronic conductance at band edges. Finally, the $ZT$ becomes smaller under strain, because the reduction in the electronic conductance overcomes the reduction in the thermal conductance.

We further remark that we have applied methods of ballistic transport to compute transport parameters, and ultimately using them to find $ZT$. At temperatures exceeding 1000 K, the phonon transport should take into consideration anharmonic phonon interactions, which would lead to a decrease in thermal conductivity at such high temperatures, rather than the saturation effect seen in these figures for the phonon thermal conductance. As a result, this would only reduce thermal conductivity and work to boost $ZT$.

\textbf{Acknowledgements} The work is supported by a Faculty Research Grant of R-144-000-257-112 of NUS, and Grant R-144-000-203-112 from Ministry of Education of Republic of Singapore, and Grant R-144-000-222-646 from NUS.


\end{document}